# Chiral Tunnelling in a Twisted Graphene Bilayer

Wen-Yu He, Zhao-Dong Chu, and Lin He*

Department of Physics, Beijing Normal University, Beijing, 100875, People's Republic of China

**The perfect transmission in graphene monolayer and the perfect reflection in Bernal graphene bilayer for electrons incident in the normal direction of a potential barrier are viewed as two incarnations of the Klein paradox. Here we show a new and unique incarnation of the Klein paradox. Owing to the different chiralities of the quasiparticles involved, the chiral fermions in twisted graphene bilayer show adjustable probability of chiral tunnelling for normal incidence: they can be changed from perfect tunnelling to partial/perfect reflection, or vice versa, by controlling either the height of the barrier or the incident energy. As well as addressing basic physics about how the chiral fermions with different chiralities tunnel through a barrier, our results provide a facile route to tune the electronic properties of the twisted graphene bilayer.**

Because graphene's two-dimensional honeycomb lattice, quasiparticles in graphene mimic Dirac fermions in quantum electrodynamics (QED) [1-8]. Therefore, this condensed-matter system is expected to help demonstrate many oddball effects predicted by QED. One example is the Klein paradox [9-11]. The chirality of the charge carriers in graphene monolayer ensures a perfect quantum tunnelling for electrons incident in the normal direction of a potential barrier [9,10]. This is viewed as a direct experimental test of the Klein's gedanken experiment [11]. In light of possible applications, the chirality suppresses back-scattering of quasiparticles and protects high charge carrier mobility of graphene despite unavoidable inhomogeneities [3,12,13]. The emergence of superlattice Dirac points in graphene superlattice, as reported very recently [14-16], is also directly related to the chiral nature of the Dirac fermions [17-20]. Owing to the different chiralities of the quasiparticles involved, the quantum tunnelling in Bernal graphene bilayer leads to the opposite effect: massive chiral fermions are always perfectly reflected for a sufficiently wide barrier for normal incidence [9]. This result implies that it may be possible to find different chiral fermions in graphene system to show "designable" tunnelling properties. In this Letter, we will demonstrate subsequently that twisted graphene bilayer is a good candidate to achieve this goal. The chiral fermions in twisted graphene bilayer shows adjustable probability of chiral tunnelling for a normal incidence. The transmission probability can be changed between 1 and 0 by controlling either the height of the barrier or the incident energy. This unique tunnelling behavior is of potential application in designing of future electronic device.

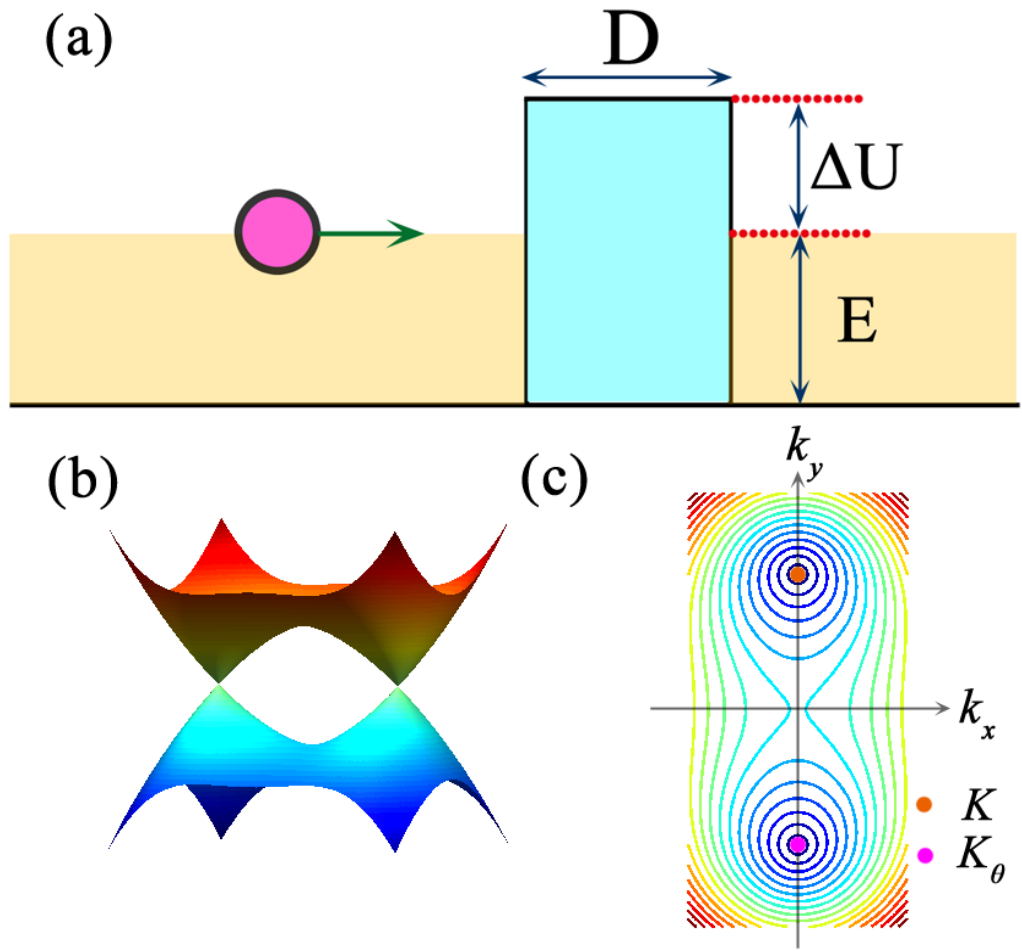

**Figure 1|** Tunnelling through a barrier in twisted graphene bilayer. **(a)** Schematic diagram of an electron coming to a potential-energy barrier of height E + ΔU and width D. E is the Fermi energy of the twisted graphene bilayer and the one-dimensional barrier is infinite along the y direction. **(b)** Electronic spectrum of the quasiparticles in twisted graphene bilayer with a finite interlayer coupling. Two saddle points form between the two Dirac cones, K and $K_\theta$. **(c)** Density plot of the energy dispersion of the twisted graphene bilayer around K and $K_\theta$. $k_x$ is the direction perpendicular to the barrier.

Figure 1(a) shows the general scheme that an chiral electron starts penetrating through a potential barrier $U(x)$, which has a rectangular shape with width of $D$ and height of $E + \Delta U$ (here $E$ is the incident energy of the electron, $\Delta U$ is the energy difference between the potential barrier and the incident energy, and $\Delta U > 0$ in our calculation). The potential barrier is infinite along the $y$ axis. The rectangular shape assumption of the barrier means that the characteristic width of the edge smearing is much smaller than the electron wavelength but much larger than the lattice constant. Such an assumption disallows sharp enough scattering to mix the two valleys in graphene and consequently we only need to consider scattering electrons from one valley [9]. This tunnelling problem was considered first in Ref. 9 for chiral electrons in graphene monolayer and Bernal graphene bilayer. This system can be divided into three distinct regions: the left of the barrier ($x < 0$), inside the barrier ($0 < x < D$), and the right of the barrier ($x > D$). If we know the wavefunctions in the three regions, then it is straightforward to solve this tunnelling problem. For a twisted graphene bilayer, the Dirac points of the two layers no longer coincide and the zero energy states occur at k = $-\Delta K/2$ and k = $\Delta K/2$ in layer 1 and 2 respectively. Here $\Delta K = 2K\sin(\theta/2)$ is the shift between the corresponding Dirac points of the twisted graphene bilayer, and $K = 4\pi/3a$ with a ~ 0.246 nm the lattice constant of the hexagonal lattice. The displaced Dirac cones of the twisted bilayer cross and two intersections of the saddle points along the two cones appear in the presence of interlayer coupling $t_\perp$ [21,22], as shown in Figure 1(b) and 1(c). The saddle points result in two low-energy van Hove singularities (VHSs) at $\pm E_v =$ $\pm 1/2(\hbar v_F \Delta K - 2t_\perp)$ in the density of states (here $v_F \sim 1.0 \times 10^6$ m/s is the Fermi velocity). The band structure of the twisted graphene bilayer was subsequently confirmed experimentally by Raman, scanning tunnelling spectroscopy, and angle-resolved photoemission spectroscopy [15,23-28].

When we consider only low-energy excitations, the effective hamiltonian of the twisted graphene bilayer can be described by [21,22,29]

$$H^{eff} = -\frac{2v_F^2}{15\tilde{t}_\perp} \begin{pmatrix} 0 & (k^*)^2 - \left(\frac{1}{2}\Delta K^*\right)^2 \\ k^2 - \left(\frac{1}{2}\Delta K\right)^2 & 0 \end{pmatrix}, \quad (1)$$

where $k = k_x + ik_y$ is the wave vector relative to the midpoint of the two Dirac points. A low-energy expansion of hamiltonian (1) around $\pm \Delta K/2$, by defining $k = q \pm \Delta K/2$, yields two Dirac hamiltonians $\pm \frac{2v_F^2 \Delta K}{15\tilde{t}_\perp} \vec{\sigma} \cdot \vec{q}$, which have identical chirality as that of graphene monolayer [29]. It indicates that the low-energy tunnelling behavior in twisted graphene bilayer should be similar to that in graphene monolayer. The energy spectrum derived from hamiltonian (1) is

$$E(k_x, k_y) = \pm \frac{2v_F^2}{15\tilde{t}_\perp} \sqrt{\left(k_x^2 - k_y^2 - \frac{1}{4}\Delta K_x^2 + \frac{1}{4}\Delta K_y^2\right)^2 + \left(2k_x k_y - \frac{1}{2}\Delta K_x \Delta K_y\right)^2}. \quad (2)$$

Unlike the case of single layer and Bernal bilayer graphene, the group velocity of wave packets in twisted bilayer graphene is not parallel to its wave vector any more. It could be determined by $\vec{v}_{k_0} = \frac{1}{\hbar}(\nabla_k E)_{k_0}$. In our calculation, we also use the rectangular shape assumption of the barrier and consequently Umklapp scattering between different valleys in graphene can be neglected. Therefore, we only consider scattering electrons from the K and $K_\theta$ cones. The velocity field of quasiparticles with various energies is shown in Fig. S1 (see Supplementary Information [30]). Inserting a trial wave function $\Psi(x,y) = \begin{pmatrix} \varphi_A(x) \\ \varphi_B(x) \end{pmatrix} e^{ik_y y}$ into equation $H\Psi = E\Psi$ with the hamiltonian (1), we obtain

$$\left(\frac{2v_F^2}{15t_\perp}\right)^2 \left[\left(k_x^2 - k_y^2 - \frac{1}{4}\Delta K_x + \frac{1}{4}\Delta K_y\right)^2 + \left(2k_x k_y - \frac{1}{2}\Delta K_x \Delta K_y\right)^2\right] = E^2. \qquad (3)$$

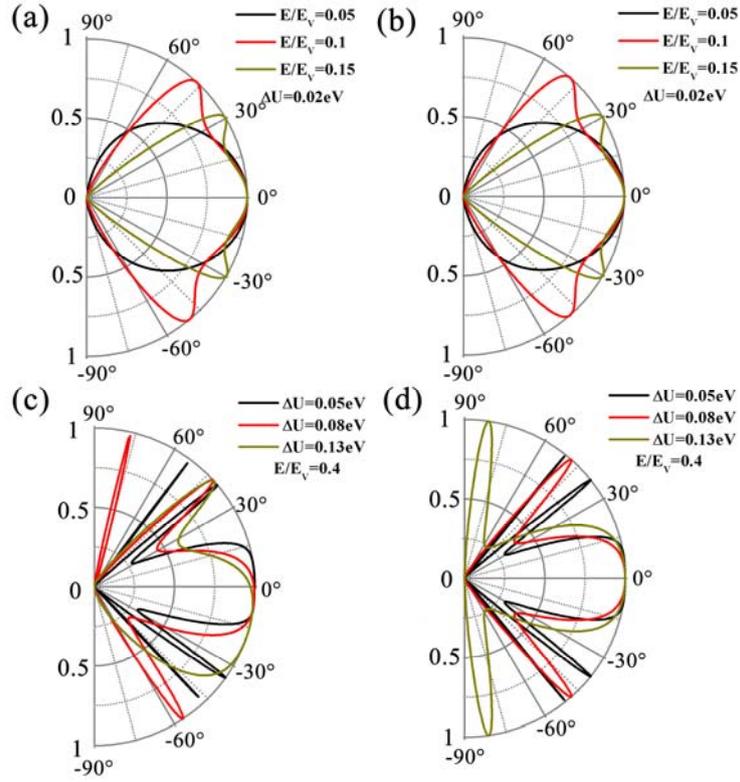

**Figure 2|** Quantum tunnelling in twisted graphene bilayer for low incident energy. Transmission probability *T* through a 100-nm wide barrier as a function of the incident angle φ for (**a**,**c**) twisted graphene bilayer and (**b**,**d**) the system described by the hamiltonian $\frac{2v_F^2 \Delta K}{15\tilde{t}_\perp}\vec{\sigma}\cdot\vec{q}$. The remaining parameters are the twist angle of the graphene bilayer θ = 3.89°, $t_\perp$ = 0.12 eV, and $E_v$ = 0.15 eV. The angular behavior of *T*(φ) in (**b**,**d**) is similar to that of graphene monolayer and the chiral fermions are always perfectly tunneling for normal incidence irrespective of the parameters of the barrier. The *T*(φ) of twisted graphene bilayer shows both similarities and differences with respect to that of graphene monolayer. For twisted graphene bilayer, the *T*(φ) is asymmetric about φ = 0 and the asymmetry increases with increasing the height of the barrier.

There are four possible solutions for a given energy. Two of them are propagating waves $\exp(\pm ik_{x1}x)$ and the other two are exponentially growing and decaying modes $\exp(\pm k_{x2}x)$. Here $k_{x1}$ and $ik_{x2}$ are wave vectors. The wave function in the three different regions of the tunnelling problem can be written in terms of incident and reflected waves. The reflection coefficient and the transmission coefficient are determined from the continuity of the wave functions and their derivatives (see Supplementary Information [30] for details of analysis and calculation).

Figure 2 shows examples of the transmission probability as a function of the incident angle $T(\varphi)$ for a twisted graphene bilayer (here we only show the transmission probability of quasiparticles in the K cone and that in the $K_\theta$ cone is mirror-symmetric about $\varphi = 0°$).

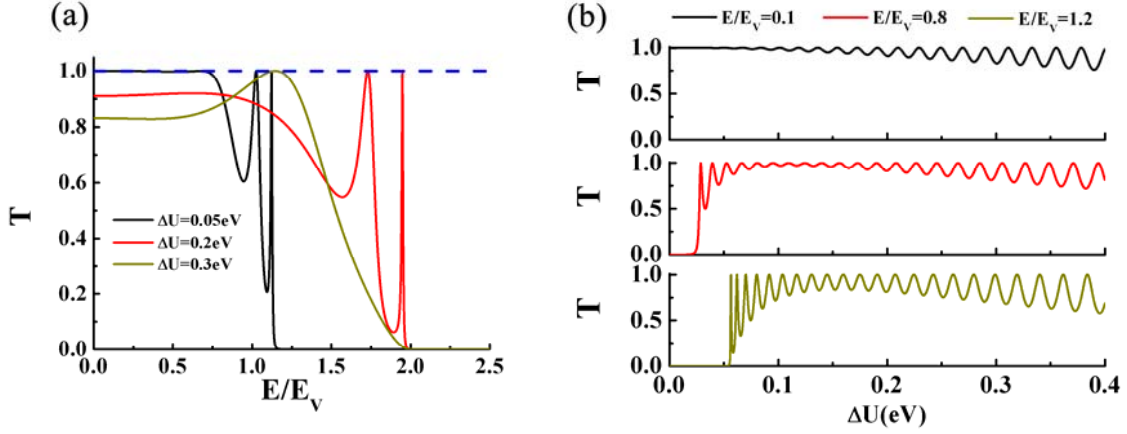

**Figure 3|** Quantum tunnelling in twisted graphene bilayer for normally incident electrons. **(a)** Transmission probability for normally incident electrons as a function of the incident energy. The curves with different colours correspond to different values of the ΔU. **(b)** Transmission probability for normally incident electrons with different incident energy as a function of the value of ΔU. In the middle and lower panels, the transmission probability is suppressed for low energy barrier and becomes an oscillating mode for higher ΔU. For a certain value of ΔU, the amplitude of the oscillations increases with the incident energy. For a fixed incident energy, both the periodicity and the amplitude of the oscillations increase with the value of ΔU.

To elucidate differences and similarities of the $T(\varphi)$ between the twisted graphene bilayer and graphene monolayer, we also calculated the same tunnelling problem of the system described by the hamiltonian $\frac{2v_F^2 \Delta K}{15\tilde{t}_\perp}\vec{\sigma}\cdot\vec{q}$ for comparison. For low incident energy and small value of ΔU (in our calculation, the potential energy is always larger than the incident energy of electrons, i.e., ΔU > 0), the angular dependence of transmission probability for the twisted graphene bilayer resembles that of graphene monolayer and the chiral fermions are perfectly or almost perfectly tunnelling for normal incidence. The differencies emerge for large incident energy and large value of ΔU. The $T(\varphi)$ is asymmetric about $\varphi = 0$ in twisted graphene bilayer and the asymmetry increases with increasing the incident energy and the height of the barrier (see Supplementary Information [30]). The most striking result of the tunnelling problem is that the transmission probability at $\varphi = 0$ depends sensitively on the incident energy and the height of the barrier, which is quite different from that of graphene monolayer and Bernal graphene bilayer.

To further understand the chiral tunnelling in twisted graphene bilayer, we calculate transmission probability for normally incident electrons as a function of the incident energy and as a function of the value of ΔU, as shown in Fig. 3(a) and 3(b) respectively. It is interesting to note that the incident energy can tune the transmission probability in twisted graphene bilayer. This unique behavior is essentially due to the different chirality or pseudospins of the quasiparticles involved. For graphene monolayer and Bernal graphene bilayer, the propagating wavefunctions can be written as $\frac{1}{\sqrt{2}}\begin{pmatrix} 1 \\ se^{i\varphi} \end{pmatrix} e^{i\vec{k}\cdot\vec{r}}$ and $\frac{1}{\sqrt{2}}\begin{pmatrix} 1 \\ se^{2i\varphi} \end{pmatrix} e^{i\vec{k}\cdot\vec{r}}$ respectively (here $s = \text{sgn}E$). The $e^{i\varphi}$ and $e^{2i\varphi}$ can be viewed as phase difference between the two components of the "spinor wavefunctions", which is independent of the incident energy. However, for the case of twisted graphene bilayer, the propagating wavefunction has the form $\frac{1}{\sqrt{2}}\begin{pmatrix} 1 \\ sq_1^+ \end{pmatrix} e^{i\vec{k}\cdot\vec{r}}$ and the term $q_1^+$ depends on the incident energy (see Supplementary Information [30]). The term $q_1^+$ of the wavefunction approaches $e^{i\varphi}$ around the Dirac points and is a good approximation of $e^{2i\varphi}$ for high-energy spectrum (the normal tunnelling becomes completely forbidden for incident energy higher than 2E$_V$, see Supplementary Information [30] for details of discussion). Therefore, the transmission probability in twisted graphene bilayer is a function of the incident energy and can be changed from perfect tunnelling to complete reflection, as shown in Fig. 3(a).

The perfect matching between an incident electron wavefunction and the corresponding wavefunction for a propagating hole inside a barrier at the barrier interface yields $T = 1$ in graphene monolayer. For Bernal graphene bilayer, the propagating electron wavefunction transforms into an evanescent hole wavefunction inside the barrier, resulting in prefect reflection for a wide barrier [9]. The chiral electrons in twisted graphene bilayer combine the two distinct behaviors of quasiparticles in graphene monolayer and Bernal graphene bilayer. Therefore, the transmission probability can be changed from perfect tunnelling to partial reflection, or vice versa, as shown in Fig. 3(b). As the amplitude of the oscillating tunnelling is amplified with increasing the value of ΔU, it is hopefully to see the transmission probability can be even switched between T = 1 and T = 0, which correspond to that the chiral electron transforms into either an propagating hole or an evanescent hole inside the barrier respectively (see Supplementary Information [30]).

In twisted graphene bilayer, the transmission probability for normal incidence with a large incident energy is zero for a small $\Delta U$, as shown in Fig. 3(b). This peculiar behavior can be attributed to the failure in creation of an electron-hole("positron") pair at the barrier interface. For a large incident energy, there exists a gap between electrons and holes in the spectrum (see Fig. S3 [30]). A small potential barrier cannot overcome the energy gap to excite holes in the classical forbidden area, so the wave vector inside the potential is imaginary, inhibiting the propagation [31,32]. Then it only shows exponentially decaying tunnelling and the transmission probability is zero for a wide barrier.

The periodicity of the oscillations, as shown in Fig. 3(b), increases with increasing the value of $\Delta U$. This effect can be explained with the help of the quantum confinement of the propagating wavefunctions inside the barrier. When the electron wavefunction perfectly matches the wavefunction for a propagating hole, the barrier is transparent. The energy interval between the nearest states of the propagating hole wavefunctions inside the barrier is proportional to height of the barrier. As a consequence, the periodicity of the transmission probability increases with the height of barrier. To further confirm the above analysis, it is helpful to consider the same tunnelling problem with different width of the barrier. The energy interval between the nearest states of the propagating hole wavefunctions inside the barrier is expected to increase linear with the inverse of the width of the barrier $D^{-1}$. Therefore, the periodicity of the oscillations should increase linear with $D^{-1}$, which is confirmed explicitly by the result shown in Fig. 4.

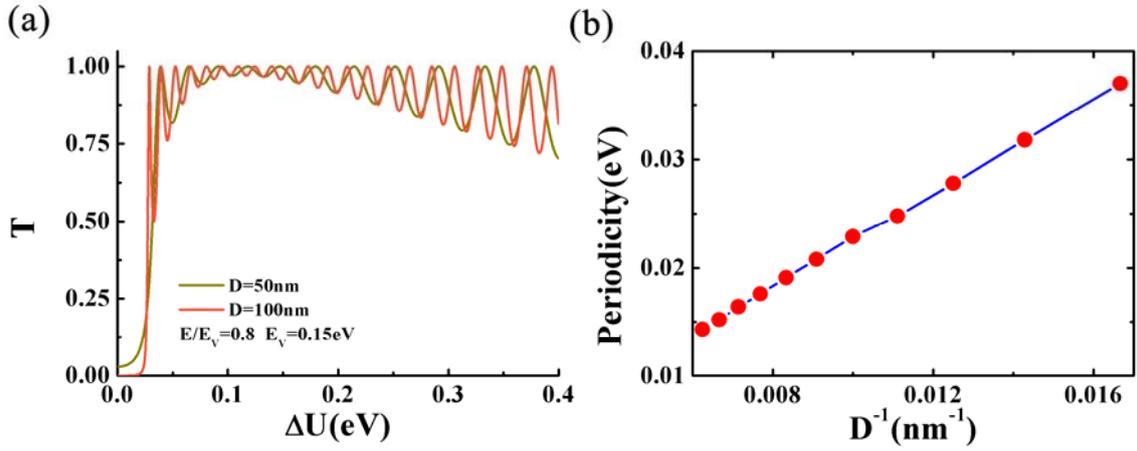

**Figure 4|** Quantum tunnelling in twisted graphene bilayer with different width of the barrier. **(a)** Transmission probability as a function of the values of the ΔU. The curves with different colours correspond to different width of the barriers. **(b)** The average periodicity of the oscillations for 0.3 eV < ΔU < 0.4 eV as a function of $D^{-1}$. Here D is the width of the barrier. The periodicity almost increases linear with $D^{-1}$.

The perfect chiral tunnelling of graphene monolayer inhibits the fabrication of standard semiconductor devices because field effect transistors made from graphene monolayer remain conducting even when switched off [2]. The ability to control the transmission of quasiparticles through a barrier in twisted graphene bilayer suggests that this effect can be used as the basis for future graphene device electronics. Experimentally, a potential barrier can be easily created by the electric field effect and the parameter of the barrier is tunable. Therefore, the predicted effect of this paper is expected to be realized in the near future.

*Email:helin@bnu.edu.cn.

**Acknowledgements**

This work was supported by the National Natural Science Foundation of China (Grant No. 11004010), the Fundamental Research Funds for the Central Universities, and the Ministry of Science and Technology of China.